RE-WEIGHTING OF SOMATOSENSORY INPUTS FROM THE FOOT AND THE ANKLE

FOR CONTROLLING POSTURE DURING QUIET STANDING

FOLLOWING TRUNK EXTENSOR MUSCLES FATIGUE

Nicolas VUILLERME and Nicolas PINSAULT

Laboratoire TIMC-IMAG, UMR UJF CNRS 5525, La Tronche, France

Address for correspondence:

Nicolas VUILLERME

Laboratoire TIMC-IMAG, UMR UJF CNRS 5525

Faculté de Médecine

38706 La Tronche cédex

France.

Tel: (33) (0) 4 76 63 74 86

Fax: (33) (0) 4 76 51 86 67

Email: nicolas.vuillerme@imag.fr

**Abstract**

The present study focused on the effects of trunk extensor muscles fatigue on postural control during quiet standing under different somatosensory conditions from the foot and the ankle. With this aim, 20 young healthy adults were asked to stand as immobile as possible in two conditions of No fatigue and Fatigue of trunk extensor muscles. In Experiment 1 (n = 10), somatosensation from the foot and the ankle was degraded by standing on a foam surface. In Experiment 2 (n = 10), somatosensation from the foot and ankle was facilitated through the increased cutaneous feedback at the foot and ankle provided by strips of athletic tape applied across both ankle joints. The centre of foot pressure displacements (CoP) were recorded using a force platform. The results showed that (1) trunk extensor muscles fatigue increased CoP displacements under normal somatosensatory conditions (Experiment 1 and Experiment 2), (2) this destabilizing effect was exacerbated when somatosensation from the foot and the ankle was degraded (Experiment 1), and (3) this destabilizing eVect was mitigated when somatosensation from the foot and the ankle was facilitated (Experiment 2). Altogether, the present Wndings evidenced re-weighting of sensory cues for controlling posture during quiet standing following trunk extensor muscles fatigue by increasing the reliance on the somatosensory inputs from the foot and the ankle. This could have implications in clinical and rehabilitative areas.

**Introduction**

Muscle fatigue represents an inevitable phenomenon for physical, professional and daily activities that the central nervous system (CNS) has to take into account. In recent years, a growing number of studies have reported increased postural sway during quiet standing with muscle fatigue localized at the lower back (Davidson et al. 2004; Madigan et al. 2006; Pline et al. 2006; Vuillerme et al. 2007). Although the exact mechanism inducing these postural impairments is rather diYcult to be determined, it is likely that an alteration of the functionality of the sensory proprioceptive and motor systems caused by trunk muscles fatiguing exercise explained these observations. Indeed, previous studies have reported that trunk muscles fatigue altered proprioceptive acuity at the ankle (Pline et al. 2005) and the torso (Taimela et al. 1999), delayed the reaction time of the muscles in response to a sudden load (Wilder et al. 1996), reduced the force-generating capacity (e.g. Ng et al. 2003; Potvin and O'Brien 2002) and increased its variability (e.g. Ng et al. 2003; Potvin and O'Brien 2002).

Interestingly, the abovementioned studies assessed the postural eVects of trunk extensor muscles fatigue under normal somatosensory conditions from the foot and ankle. Considering the important role of foot and ankle somatosensory inputs in the regulation of postural sway during quiet standing (e.g. Kavounoudias et al. 2001; Meyer et al. 2004), the present study was thus designed to assess the effects of trunk extensor muscles fatigue on postural control during quiet standing under diVerent conditions of availability and/or accuracy and/or reliability of somatosensory inputs from the foot and the ankle. It was hypothesized that (1) trunk extensor muscles fatigue would increase postural sway during quiet standing and (2) this effect would depend on the availability, accuracy and/or the reliability of the somatosensory information at the foot and ankle. Specifically, we expected that an alteration and a facilitation of somatosensory inputs from the foot and the ankle would exacerbate and mitigate the destabilizing effect of trunk muscles fatigue, respectively.

## Methods

Ten young university students (age: $26.0 \pm 5.6$ years; body weight: $73.7 \pm 8.9$ kg; height: $180.2 \pm 6.4$ cm; mean § S.D.) participated in Experiment 1. Ten other young university students (age: $24.5 \pm 4.2$ years; body weight: $74.3 \pm 7.4$ kg; height: $178.5 \pm 5.2$ cm) took part in Experiment 2. They gave their informed consent to the experimental procedure as required by the Helsinki declaration (1964) and the local Ethics Committee and were naive as to the purpose of the experiment. None of the subjects presented any history of motor problem, neurological disease or vestibular impairment. With their eyes closed, subjects stood barefoot on a force platform in a natural position (feet abducted at $30°$, heels separated by 3 cm), their arms hanging loosely by their sides and were asked to sway as little as possible. The force platform (Equi+, model PF01), which constituted of an aluminium plate (80 cm each side) lying on three uniaxial load cells, was used to measure the displacements of the centre of foot pressure (CoP). Signals from the force platform were sampled at 64 Hz, amplified and converted from analogue to digital form.

In Experiment 1, the postural task was performed on two Firm and Foam support surface conditions. The force platform served as the Firm support surface. In the Foam condition, a 2-cm thick foam support surface, altering the quality and/or quantity of somatosensory information at the foot sole and the ankle, was placed under the subjects' feet (Vuillerme et al. 2001a, 2005; Isableu and Vuillerme 2006).

In Experiment 2, the postural task was performed on a Firm support surface in two conditions of No tactile stimulation and Tactile stimulation of the foot and ankle. The No tactile stimulation condition served as a control condition. In the Tactile stimulation condition, two pieces of 5-cm wide strips of athletic tape were applied in a distal-proximal direction directly to the skin in front of and behind the subject's talocrural joints (Simoneau et al. 1997). The first strip, starting approximately 10 cm proximal to the ankle joint line and ending 5 cm distal to the ankle joint line, was positioned directly on the skin over the anterior aspect of the ankle joint. The second strip was used posteriorly over the Achilles tendon and calcaneus. These strips of tape, used to selectively provide cutaneous sensory feedback around both ankles without the added mechanical constriction and mechanical pressure on subcutaneous structures associated with the application of ankle taping as used in athletic events, have previously been shown to improve ankle proprioceptive acuity in young healthy subjects (Simoneau et al. 1997).

For both the Experiment 1 and Experiment 2, this experimental procedure was executed the same day before (No fatigue condition) and immediately after a designated fatiguing exercise for trunk extensor muscle (Fatigue condition). The muscular fatigue was induced until maximal exhaustion with trunk repetitive extensions as recently described by Vuillerme et al. (2007). Subjects lay prone on a bench with the upper body unsupported in the horizontal plane. The lower extremities were secured to the bench with straps at the hips, knees and ankles. During the test, arms were held crossed the chest. Subjects were instructed to raise their upper body to a horizontal position and then lower it back down as many times as possible following the beat of a metronome (40 beats/min). Verbal encouragement was given to ensure that subjects worked maximally. The fatigue level was reached when subjects were no more able to complete the trunk extension exercise. Immediately on the cessation of exercise, the subjective exertion level was assessed through the Borg CR-10 scale (Borg 1990). Subjects rated their perceived fatigue in the trunk extensor muscles as almost "extremely strong" (mean Borg ratings of 8.4 and 8.8, for Experiment 1 and Experiment 2, respectively). The recovery process after fatigue procedures is often considered as a limitation for all fatigue experiments. In the present study, to ensure that balance measurement in the Fatigue condition was obtained in a genuine fatigued state, various rules were respected (Vuillerme et al. 2001b, 2002a, b, 2005, 2006, 2007; Vuillerme and Demetz 2007; Vuillerme and Nougier 2003). (1) The fatiguing exercise took place beside the force platform, so that there was a short time-lag between the exercise-induced fatiguing activity and the balance measurements and (2) the fatiguing exercise was repeated prior to each trial.

For each somatosensory condition (the two Firm and Foam support surface conditions and the two No tactile and Tactile stimulation of the foot and ankle conditions, for Experiment 1 and Experiment 2, respectively) and each condition of No fatigue and Fatigue of the trunk extensor muscles (for Experiment 1 and Experiment 2), subjects performed three 30-s trials, for a total of 12 trials. For each experiment, the order of presentation of the two somatosensory conditions from the foot and the ankle was randomized over subjects.

Centre of foot pressure displacements were processed through a space–time domain analysis including the calculation of the surface area (mm²) covered by the trajectory of the CoP with a 90% conWdence interval (Tagaki et al. 1985). This dependent variable provides a measure of the CoP spatial variability.

The means of the three trials performed in each of experimental condition were used for statistical analyses. A 2 Fatigues (No fatigue *vs.* Fatigue) × 2 Support surfaces (Firm *vs.* Foam) analyses of variances (ANOVA) with repeated measures of both factor was applied to data obtained in Experiment 1. A 2 Fatigues (No fatigue *vs.* Fatigue) × 2 Tactile stimulations (No tactile stimulation *vs.* Tactile stimulation) ANOVA with repeated measures of both factors was applied to data obtained in Experiment 2. Post hoc analyses (*Newman-Keuls*) were performed whenever necessary. Level of significance was set at 0.05.

**Results**

Experiment 1

Analysis of the surface area covered by the trajectory of the CoP showed a significant interaction of Fatigue × Support surface [$F(1,9) = 9.37$, $P < 0.05$]. As illustrated in Fig. 1, the decomposition of this interaction into its simple main effects indicated that (1) the Fatigue condition yielded larger CoP surface area relative to the No fatigue condition in the Firm condition ($P < 0.01$) and (2) this effect was more accentuated in the Foam condition ($P < 0.001$). The ANOVA also showed a significant main effect of Support surface [$F(1,9) = 33.44$, $P < 0.001$], yielding an increased surface area in the Foam relative to the Firm condition.

Experiment 2

Analysis of the surface area covered by the trajectory of the CoP showed a significant interaction of Fatigue × Tactile stimulation [$F(1,9) = 5.69$, $P < 0.05$]. As illustrated in Fig. 2, the decomposition of this interaction into its simple main effects indicated that (1) the Fatigue condition yielded larger CoP surface area relative to the No fatigue condition in the No tactile stimulation condition ($P < 0.001$) and (2) this effect was mitigated in the Tactile stimulation condition ($P < 0.05$). The ANOVAs also showed main effects of Fatigue [$F(1,9) = 31.11$, $P < 0.001$] and Tactile stimulation [$F(1,9) = 8.03$, $P < 0.05$], yielding increased surface area in the Fatigue relative to the No fatigue condition and decreased surface area in the Tactile stimulation relative to the No Tactile stimulation condition, respectively.

**Discussion**

The present study focused on the effects of trunk extensor muscles fatigue on postural control during quiet standing under different somatosensory conditions from the foot and the ankle. In normal somatosensory conditions from the foot and ankle, our results showed that trunk extensor muscles fatigue impaired postural control during quiet standing, as indicated by a wider surface area covered by the CoP trajectory observed in the Fatigue than No Fatigue condition (Figs. 1, 2, left part). This result confirms our hypothesis 1, in accordance with previous reports (Davidson et al. 2004; Madigan et al. 2006; Pline et al. 2006; Vuillerme et al. 2007).

Beyond these well-established results, our results further evidenced that the effects of trunk extensor muscles fatigue during quiet standing depended on the availability, accuracy and/or reliability of somatosensory inputs from the foot and the ankle, confirming our hypothesis 2.

In Experiment 1 (n = 10), somatosensation from the foot and ankle was degraded by standing on a foam surface (Vuillerme et al. 2001a, 2005; Isableu and Vuillerme 2006). The observation of a signiWcant interaction Fatigue × Support surface (Fig. 1) showed that this destabilizing effect of trunk extensor muscles fatigue was exacerbated when somatosensation from the foot soles and ankles was degraded.

In Experiment 2 (n = 10), somatosensation from the foot and ankle was facilitated through the increased cutaneous feedback at the foot and ankle provided by strips of athletic tape applied across both ankle joints (Simoneau et al. 1997). The observation of a signiWcant interaction Fatigue × Tactile stimulation (Fig. 2) showed that the destabilizing effect of trunk extensor muscles fatigue was mitigated when somatosensation from the foot soles and ankles was facilitated. This result suggests that the CNS was able to integrate the afferent input from cutaneous mechanoreceptors in the foot and shank (stimulated by the pressure and traction of the material on the skin) to limit the postural destabilization induced by trunk extensor muscles fatigue.

Altogether, results of Experiment 1 and Experiment 2 evidenced an increased reliance on somatosensory inputs from the foot soles and ankles for controlling posture during quiet standing following trunk extensor muscles fatigue. These findings could be attributable to sensory reweighting hypothesis (e.g. Horak and Macpherson 1996; Oie et al. 2002; Peterka 2002; Peterka and Loughlin 2004; Vuillerme et al. 2001b, 2002a, 2005, 2006; Vuillerme and Demetz 2007; Vuillerme and Nougier 2003), whereby the CNS dynamically and selectively adjusts the relative contributions of sensory inputs (i.e. the sensory weights) to maintain upright stance depending not only on the sensory environment, but also on the neuromuscular constraints acting on the subject. Finally, the results of the present study, stressing the importance of accurate and reliable somatosensory inputs from foot and ankle, could have implications in ergonomical, clinical and rehabilitative areas.

**Acknowledgements**

The authors would like to thank subject volunteers. This research was supported by the MENRT. Special thanks also are extended to Claire H. and P. Hompideup for various contributions.

**References**

Borg G (1990) Psychological scaling with applications in physical work and the perception of exertion. Scand J Work Environ Health 16:55–58

Davidson BS, Madigan ML, Nussbaum MA (2004) Effects of lumbar extensor fatigue and fatigue rate on postural sway. Eur J Appl Physiol Occup Physiol 93:183–189

Horak FB, Macpherson JM (1996) Postural orientation and equilibrium. In: Rowell LB, Shepard JT (eds) Handbook of physiology. Exercise: regulation and integration of multiple systems. Oxford University Press, Oxford, pp 255–292

Isableu I, Vuillerme N (2006) Differential integration of kinaesthetic signals to postural control. Exp Brain Res 174:763–768

Kavounoudias A, Roll R, Roll JP (2001) Foot sole and ankle muscle inputs contribute jointly to human erect posture regulation. J Physiol 532:869–878

Madigan ML, Davidson BS, Nussbaum MA (2006) Postural sway and joint kinematics during quiet standing are affected by lumbar extensor fatigue. Hum Mov Sci 25:788–799

Meyer PF, Oddsson LIE, De Luca CJ (2004) The role of plantar cutaneous sensation in unperturbed stance. Exp Brain Res 156:505–512

Ng JKF, Parnianpour M, Richardson CA, Kippers V (2003) Effect of fatigue on torque output and electromyographic measures of trunk muscles during isometric axial rotation. Arch Phys Med Rehabil 84:374–381

Oie KS, Kiemel T, Jeka JJ (2002) Multisensory fusion: simultaneous re-weighting of vision and touch for the control of human posture. Brain Res Cogn Brain Res 14:164–176

Peterka RJ (2002) Sensorimotor integration in human postural control. J Neurophysiol 88:1097–1118

Peterka RJ, Loughlin PJ (2004) Dynamic regulation of sensorimotor integration in human postural control. J Neurophysiol 91:410–423

Pline KM, Madigan ML, Nussbaum MA, Grange RW (2005) Lumbar extensor fatigue and circumferential ankle pressure impair ankle joint motion sense. Neurosci Lett 390:9–14

Pline KM, Madigan ML, Nussbaum MA (2006) Influence of fatigue time and level on increases in postural sway. Ergonomics 49:1639–1648

Potvin JR, O'Brien PR (2002) Trunk muscle co-contraction increases during fatiguing, isometric, lateral bend exertions. Possible implications for spine stability. Spine 23:774–780 (discussion 781)

Simoneau GG, Degner RM, Kramper CA, Kittleson KH (1997) Changes in ankle joint proprioception resulting from strips of athletic tape applied over the skin. J Athl Train 32:141–147

Tagaki A, Fujimura E, Suehiro S (1985) A new method of statokinesigram area measurement. Application of a statistically calculated ellipse. In: Igarashi M, Black O (eds) Vestibular and visual control on posture and locomotor equilibrium. Karger, Bâle, pp 74–79

Taimela S, Kankaanpaa M, Luoto S (1999) The eVect of lumbar fatigue on the ability to sense a change in lumbar position. A controlled study. Spine 24:1322–1327

Vuillerme N, Nougier V (2003) EVect of light Wnger touch on postural sway after lower-limb muscular fatigue. Arch Phys Med Rehabil 84:1560–1563

Vuillerme N, Demetz S (2007) Do ankle foot orthoses modify postural control during bipedal quiet standing following a localized fatigue at the ankle muscles? Int J Sports Med 28:243–246

Vuillerme N, Marin L, Debû B (2001a) Assessment of static postural control in teenagers with Down Syndrome. Adapt Phys Act Q 18:417–433

Vuillerme N, Nougier V Prieur JM (2001b) Can vision compensate for a lower limbs muscular fatigue for controlling posture in humans? Neurosci Lett 308:103–108

Vuillerme N, Danion F, Forestier N, Nougier V (2002a) Postural sway under muscle vibration and muscle fatigue in humans. Neurosci Lett 333:131–135

Vuillerme N, Forestier N, Nougier V (2002b) Attentional demands and postural sway: the eVect of the calf muscles fatigue. Med Sci Sports Exerc 34:1607–1612

Vuillerme N, Pinsault N, Vaillant J (2005) Postural control during quiet standing following cervical muscular fatigue: effects of changes in sensory inputs. Neurosci Lett 378:135–139

Vuillerme N, Burdet C, Isableu B, Demetz S (2006) The magnitude of the eVect of calf muscles fatigue on postural control during bipedal quiet standing with vision depends on the eye-visual target distance. Gait Posture 24:169–172

Vuillerme N, Anziani B, Rougier P (2007) Trunk extensor muscles fatigue aVects undisturbed postural control mechanisms in young healthy adults. Clin Biomech 22:489–494

Wilder DG, Aleksiev AR, Magnusson ML, Pope MH, Spratt K, Goel VK (1996) Muscular response to sudden load—a tool to evaluate fatigue and rehabilitation. Spine 21:2628–2639

**Figure captions**

**Figure 1.** Mean and standard error of surface area covered by the trajectory of the CoP obtained for the two Firm and Foam support surface conditions and the two conditions of No fatigue and Fatigue of trunk extensor muscles. The two conditions of No fatigue and Fatigue are presented with different symbols: No fatigue (white bars) and Fatigue (black bars). The significant P-values for comparisons between the No fatigue and Fatigue conditions are also reported (**$P < 0.01$; ***$P < 0.001$).

**Figure 2.** Mean and standard error of surface area covered by the trajectory of the CoP obtained for the two No tactile stimulation and Tactile stimulation of the foot and ankle conditions, the two conditions of No fatigue and Fatigue of trunk extensor muscles. The two conditions of No fatigue and Fatigue are presented with different symbols: No fatigue (white bars) and Fatigue (black bars). The signiWcant P-values for comparisons between the No fatigue and Fatigue conditions are also reported (*$P < 0.05$; ***$P < 0.001$)

**Figure 1.**

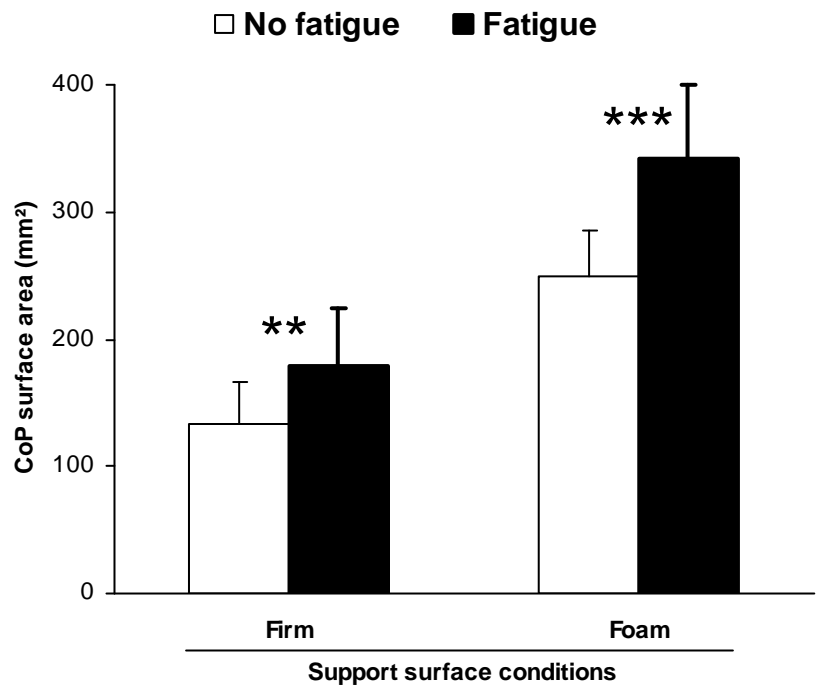

**Figure 2.**

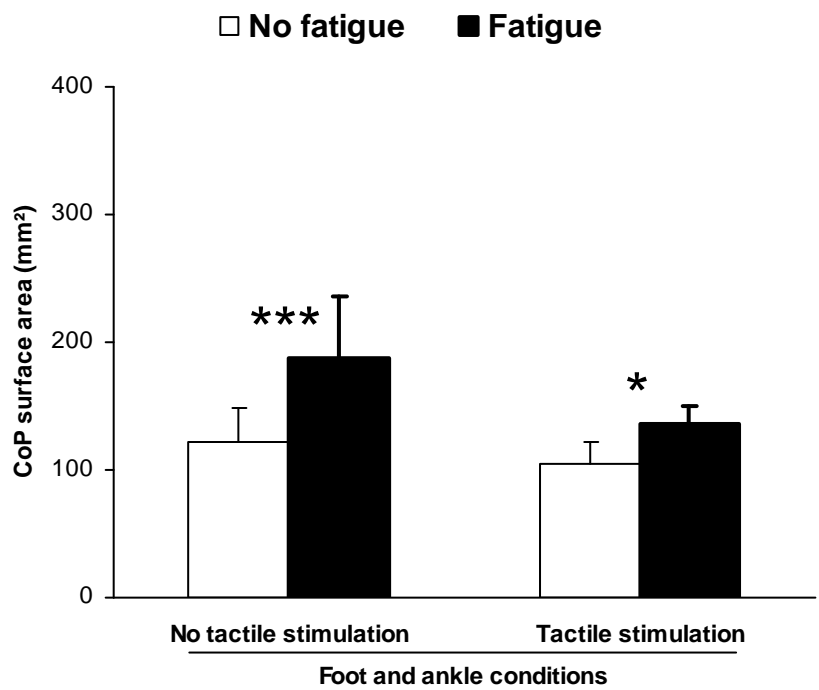